\documentstyle[12pt]{article}

\begin{document}

\title{Correlation Dynamics of Quantum Fields
and Black Hole Information Paradox
\thanks{ Lectures given at the International School of Astrophysics
``D. Chalonge": {\it String Gravity and Physics at the Planck Energy Scale},
Erice, Sept. 1995. Proceedings edited by  N. Sanchez,
(World Scientific, Singapore, 1996)}}
\author{B. L. Hu\thanks{ Email: hu@umdhep.umd.edu}\\
{\small Department of Physics, University of Maryland,
College Park, MD 20742, USA}}
\date{\small (umdpp 96-51, November 15, 1995)}
\maketitle
\begin{abstract}
In recent years  a statistical mechanics description of
particles, fields and spacetime based on the concept of quantum open
systems and the influence functional formalism has been introduced.
It reproduces in full the
established theory of quantum fields in curved spacetime and contains also
a microscopic description of their statistical properties, such as noise,
fluctuations, decoherence, and dissipation. This new framework
allows one to explore the quantum statistical properties of spacetime
at the interface between the semiclassical and quantum gravity regimes,
as well as important non-equilibrium processes
in the early universe and black holes,
such as particle creation, entropy generation, galaxy formation,
Hawking radiation, gravitational collapse, backreaction
and the black hole end-state and information lost issues.
Here we give a summary of the theory of correlation dynamics of quantum fields
and describe how  this conceptual scheme coupled with scaling behavior
near the infrared limit can shed light on the black hole information paradox.

\end{abstract}
\maketitle

\newpage

\section{Stochastic Theory of Quantum Fields and Spacetime}

I gave two lectures at this School. Since they contain material
already published, I do not want to repeat them here. Instead, I will mention
some on-going projects in our research program and discuss some ideas on how
the correlations of quantum fields can be used to address the black hole
information paradox issue. The abstracts of these two lectures, and
relevant references (where the background material and bibliography
can be found) are given below. (For a more comprehensive recent review,
see \cite{Banff})

\subsection*{Lecture 1. Stochastic Analysis of Particles and Fields: The
Effective
Action and the Influence Functional Methods}

We first summarize the in-out (Schwinger-DeWitt) effective action formalism
\cite{SchDeW} in quantum field theory in curved spacetime \cite{BirDav}
and point out the need for generalization to the in-in \cite{ctp}
(Schwinger-Keldysh, or closed-time-path CTP)
\cite{CH87,CH88,CH89} formalism to treat particle creation
and backreaction problems in semiclassical gravity \cite{HuPhysica}.
We then discuss the statistical
mechanics of quantum open systems by way of the Brownian motion model,
using the influence functional (IF) \cite{if} formalism of Feynman-Vernon
\cite{HPZ1,HPZ2}. By viewing the
subjects of interest in semiclassical gravity as open systems, where the
classical geometry is treated as the system and the quantum matter field as
the environment, and using the close relation of the CTP and
the IF methods, we can examine both the quantum and the
statistical mechanical attributes of the processes involved.
We derive an expression for the CTP effective action or the IF in terms of the
Bogolubov coefficients relating the second quantized operators between
two Fock spaces of the field theory \cite{HM2,nfsg},
thus connecting back with the established
theories of quantum fields in curved spacetimes.
As example, we show how one can derive the Hawking-Unruh radiations
\cite{Haw75,Unr76}
from a non-equilibrium statistical mechanics (NESM) viewpoint \cite{HM2}.
These effects are understood as resulting from the scaling or
amplification of quantum noise between observers in
different kinematic or dynamical states \cite{HuEdmonton,Dalian}.
Viewing these effects in this
light would enable one to consider fully non-equilibrium processes
not easily approachable by conventional (e.g., geometric or thermodynamics)
means.

\subsection*{Lecture 2. Einstein-Langevin Equation in Semiclassical Gravity:
 Backreaction as Fluctuation-Dissipation Relation}

The influence functional method in NESM (which includes the CTP effective
action
method in quantum field theory) explicates the meaning and  the
interrelation \cite{HuTsukuba,Banff}
of the many quantum processes involved in the backreaction problem, such as
particle creation, noise, fluctuations \cite{nfsg},
decoherence \cite{conhis,envdec} and dissipation \cite{HuPhysica}.
The backreaction of created particles and their fluctuations is described
by an equation of motion derivable from the influence action in the form of
an Einstein-Langevin equation \cite{HM3}.
It contains a dissipative term for the dynamics
of spacetime and a noise term related to the fluctuations of particle creation
in the matter field. As examples, we first study
the case of a free quantum scalar field in a spatially flat Friedmann-
Robertson-Walker universe and derive the Einstein-Langevin equations
for the scale factor for these semiclassical cosmologies \cite{nfsg}. Then
using the well-studied model of a quantum scalar field in a Bianchi
Type-I universe we illustrate how this Langevin equation and the noise
term are derived and show how the creation of particles and the dissipation
of anisotropy during the expansion of the very early universe can be understood
as the manifestation of a fluctuation-dissipation relation \cite{fdrsc}.
This theorem, which exists under very general conditions for dissipations
in the dynamics of a system, and the noise and fluctuations in its
environment, embodies the backreaction effect of matter fields on the
spacetime.

Our approach based on statistical field theory extends the conventional
theory of semiclassical gravity based on a semiclassical Einstein equation
with a source given by the expectation value of the energy-momentum
tensor, to that based on a Langevin-type equation, where the dynamics and
fluctuations of spacetime are driven by
quantum fluctuations of the matter field. This generalized framework is
useful for the investigation of quantum
processes in the early universe involving fluctuations, vacuum stability
and phase transition phenomena and the non-equilibrium statistical mechanics
of black holes. It is also essential to an understanding of the cross-over
behavior between quantum gravity and general relativity.

%thus making it possible to probe into the statistical properties of
%quantum fields like noise, fluctuations, entropy, decoherence and dissipation.
%Recognition of the stochastic nature of semiclassical gravity
%is an essential step towards the investigation of the behavior of
%%fluctuations,
%instability and phase transition processes associated with the crossover
%to quantum gravity.

\subsection*{Related problems }

Let me also mention some related problems currently under investigation:\\

\noindent 1) Metric Fluctuations in Semiclassical Gravity
\cite{HM3,fdrsc,CamVer}\\
2) Quantum Fluctuations and Structure Formation in the early universe
\cite{HuBelgium,nfsg,qfsf}\\
3) Correlation, Decoherence, Dissipation and Noise of Quantum Fields
\cite{dch,cddn}\\
4) Stochastic Theory of Accelerated Detectors \cite{RHA,HRKM}\\
5) Backreaction of Unruh Radiation on an Accelerated Detector and a Moving
Mirror \cite{RH,JHR}\\
6) Fluctuation-Dissipation Relation for a Radiating Black Hole \cite{HRS}\\

The first round of attack on Problems 1) to 4) have been completed recently.
Problems 5) and 6) are in progress. In the following
I will briefly discuss how correlation dynamics in quantum fields might play
a role in the black hole information paradox problem.
Sec. 2 contains a summary from \cite{cddn}. Sec. 3 contains entirely
new material. The reader should nevertheless be warned that this
investigation is still in a preliminary stage, and the ideas are
largely speculative. Also since this is more in the nature of a progress
report than a review, I have not attempted to include a complete bibliography.
Other related work can be found from the reference lists of the quoted papers.

\section{Statistical Mechanics of an Interacting Quantum Field}

We begin with an analysis of the statistical mechanics of an interacting
quantum field. This familiar subject which one learns in the first lessons
of quantum field theory is surprisingly rich in its
statistical mechanics content, much like the role an ordinary box of gas
molecules plays in Boltzmann's sophisticated theory of kinetics and
dissipation.

\subsection{Correlation, noise, and decoherence in quantum fields}

In two recent papers \cite{dch,cddn} Calzetta and I
studied the statistical mechanical properties of
interacting quantum fields in terms of the {\it dynamics of the correlation
functions}. %The main theme can be summarized as follows:
We started from the thesis that the full dynamics of an interacting
quantum field may be described by means of the Dyson- Schwinger equations
governing the infinite hierarchy of Wightman functions which measure the
correlations of the field. We showed how
this hierarchy of equations can be obtained from the variation of the infinite
particle irreducible, or {\it `master' effective action} (MEA).
Truncation of this hierarchy
gives rise to a quantum subdynamics governing a finite number of  correlation
functions (which constitute the `system'), and expression of the higher
order correlation functions (which constitute the `environment') in terms
of the lower-order ones by functional relations (`slaving' or `factorization')
induces {\it dissipation} in the dynamics of the subsystem driven
by the stochastic
fluctuations of the environment, which we call the {\it `correlation  noises'}.
These two aspects are related by the  fluctuation-dissipation relation.
This is the quantum field equivalent of the BBGKY hierarchy in
Boltzmann's theory. Any subsystem involving a finite
number of correlation functions defines an effective theory,
which is, by this reasoning, intrinsically dissipative.
The relation of loop expansion and correlation
order is expounded. We see that ordinary quantum field theory which involves
only the mean field and a two-point function, or any finite-loop effective
action in a perturbative theory are, by nature, effective theories
which possess these properties.
Histories defined by lower-order correlation functions can be decohered by the
noises from the higher order functions and acquire classical stochastic
attributes.  We think this scheme invoking the correlation order is a
natural way to describe the quantum to classical transition for a closed
system as it avoids {\it ad hoc} stipulation of the system-environment split.
It is through decoherence that the subsystem variables become classical
and the subdynamics becomes stochastic.\\

Our viewpoint here is motivated by classical kinetic theory.
In the dynamics of a dilute gas \cite{Akhiezer,Balescu,Spohn} the exact
Newton's or Hamilton's equations for the
evolution of a many body system may be translated into a Liouville
equation for the distribution function or the BBGKY hierarchy for the
sequence of partial (n- particle) distributions. This reformulation is
only formal, which involves no loss of information or predictability.
Physical  description of the dynamics comes from  truncating the
BBGKY hierarchy and introducing a factorization condition like the molecular
chaos assumption, where the higher order
distributions are substituted by functionals of the lower order.
Constructed perturbatively, this effective theory follows
only approximately the actual dynamics.
Moreover, these functionals embody some relevant boundary conditions (such as
the `weakening of correlations' hypothesis \cite{Akhiezer}), which make
them noninvariant upon time reversal. This is how dissipation in the explicitly
irreversible Boltzmann's equation appears.
\footnote{On closer examination, it is seen that
the one- particle distribution function itself describes only the mean
number of particles within a certain location in phase space; the actual
number is also subject to fluctuations. From the average size of the
equilibrium
fluctuations, which can be determined from Einstein's formula, and the
dissipative element of the dynamics, which is contained in the collision
integral, it is possible to compute the stochastic driving force consistent
with the fluctuation- dissipation relation near equilibrium \cite{KacLogan}}.\\

{\it Correlation dynamics described by the Dyson-Schwinger hierarchy
derived from the master effective action: Truncation and Factorization}\\

We want to describe a quantum field in terms of the mean field and the
(infinite number of) correlation functions.
Here, different from the conventional treatment, we view
the  `mean' field not as the actual expectation value of the field,
but rather as representing the local value of the field within one particular
history. Quantum evolution encompasses the coherent superposition
of all possible histories \cite{conhis} and these quantities are subject to
fluctuations. This naturally introduces into the theory
stochastic elements, which has hitherto been ignored in the usual description
of quantum field theory. The theory may be enlarged by including some
correlation functions as independent variables along with the `mean'
field, which themselves are subject to fluctuations.

Our starting point is the well-known fact
that the set of all Wightman functions (time ordered products of field
operators) determines completely the quantum state of a field \cite{Haag}.
Instead of following the evolution of the field in any of the conventional
representations (Schr\"odinger, Heisenberg or Dirac's), we
focus on the dynamics of the full hierarchy of Wightman functions. To this
end it is convenient to adopt Schwinger's ``closed time-path'' techniques
\cite{ctp}, and consider time ordered Green functions as a subset of all
Green functions path- ordered along a closed time loop. The
dynamics of this larger set is described by the Dyson - Schwinger equations.

We first showed that the Dyson- Schwinger hierarchy may be  obtained via the
variational principle from a functional which we call the `Master Effective
Action' (MEA). This is a formal action functional where each Wightman function
enters as an independent variable.
We then showed that any field theory based on a finite number of (mean
field plus ) correlation functions can be viewed as a subdynamics of the
Dyson- Schwinger hierarchy. The specification
of a subdynamics involves two steps:
First, the hierarchy is {\it truncated} at a certain order.
A finite set of variables,  say, the lowest nth order correlation functions,
is identified to be the `relevant' \cite{projop}
variables, which constitute the subsystem. Second, the remaining
`irrelevant' or `environment' variables, say, the n+1 to $\infty$ order
correlation functions are {\it slaved} to the former.
Slaving (or `factorization' in the Boltzmann theory)
means that irrelevant variables are substituted by set
functionals of the relevant variables.  The process of extraction of a
subdynamics from the Dyson- Schwinger hierarchy has a correlate at the level
of the effective action,
where the MEA is truncated to a functional of a finite number of variables.
The finite effective actions so obtained (the influence action \cite{if})
are generally nonlocal and complex, which is what gives rise to the noise
and dissipation in the subdynamics.
Moreover, since the slaving process generally involves the choice of an
arrow of time, it leads to irreversibility in the cloak of dissipation in the
subdynamics \cite{projop}.\\

{\it Decoherence of correlation history and correlation noise;
fluctuations and dissipation}\\

Under realistic conditions, one may not be as much
concerned with the full quantum evolution of the field as with the
development of `classical' theories where fields are described as c-numbers,
plus perhaps a small number of correlation functions to keep track of
fluctuations. These classical theories represent the physically
observable dynamics after the process of decoherence \cite{conhis,envdec}
has destroyed or diminished the coherence of the field. In any case, no
actual observation could disclose the infinite number of degrees of freedom
of the quantum field, and therefore any conceivable observational situation
may be described in the language of a suitable complex `classical'
theory in this sense.

Decoherence is brought by the effect of a coarse-grained environment
(or `irrelevant' sector) on the system (or the `relevant' sector).
In simple models, this split is imposed by hand, as when some of the fields,
or the field values within a certain region of spacetime, are chosen as
relevant. Here, we will follow the approach of our earlier work on
the decoherence of correlation histories \cite{dch}. There is no need
to select {\it a priori} a relevant sector within the theory.
Instead, we shall seek a natural criterium for successive truncations in
the hierarchy of correlations, the degree of truncation depending on the
stipulated accuracy of measurement which can be carried out on the system.
In this framework, decoherence occurs as a consequence of the fluctuations
in the higher order correlations and results in a classical dissipative
dynamics of the lower order correlations.

Here, we adopt the consistent histories formulation of quantum mechanics
\cite{conhis} for the study of the quantum to classical transition problem.
We consider the full evolution of the field described by the
Dyson- Schwinger hierarchy as a fine- grained history while histories
where only a
finite number of Wightman functions are freely specified (with all others
slaved to them) are therefore coarse- grained. We have shown that the
finite effective actions obtained for the subsystems of lower-order
correlations are related to the decoherence functional between two such
histories of correlations \cite{dch}, its acquiring an imaginary part
signifies the existence of noise which facilitates decoherence.
Thus decoherence of correlation histories is a necessary condition for the
relevance of the c-number theory as a description of observable phenomena.
It can be seen that if the c-number theory which emerges from the quantum
subdynamics  is dissipative, then it must also be stochastic.
\footnote{ Because the fundamental variables are quantum in nature,
and therefore subject to fluctuations, a classical, dissipative dynamics
would demand the accompaniment of stochastic sources (in agreement with the
`fluctuation - dissipation theorems', for, otherwise, the theory would
permit unphysical phenomena as the damping away of zero - point fluctuations.)
Of course, these uncontrollable fluctuations may be seen at
the origin of many phenomena where structure seems to spring `out of nothing',
such as the nucleation of inhomogeneous true vacuum bubbles in a supercooled
false vacuum, or the development of inhomogeneities out of a homogeneous early
universe.} From our correlation history viewpoint, the stochasticity is in fact
not confined to the field distributions-- the correlation functions would
become
stochastic as well \cite{KacLogan}.

Following Feynman and Vernon \cite{if} in their illustration of
how noise can be defined from the influence functional, we can
relate the imaginary part of the finite effective actions
describing the truncated correlations to the auto-correlation
of the stochastic sources, i.e., correlation noises, which  drive the
c-number fields and their correlation functions via the  Langevin- type
equations. From the properties of the complete (unitary) field theory
which constitutes the  closed (untruncated) system, one can show that
the imaginary part of the effective action is related  to the nonlocal
part of the real part of the effective action which depicts dissipation.
This is where the fluctuation- dissipation theorem for
non-equilibrium systems originates \cite{fdrsc}.
%Thus noise is necessarily linked with dissipation, and vice versa.

We thus see once again the intimate connection amongst the three aspects
of the theory, decoherence, dissipation, and fluctuations
\cite{HuTsukuba,GelHar2}, now manifesting in the hierarchy of correlations
which defines the subsystems.\\

\section{Correlations in a Quantum Field and Black Hole Information Paradox}

We view the black hole and the quantum field with Hawking radiation
as a closed system. Even though the quantum field might be assumed to be
free in the beginning, interaction still exists in its coupling with
the black hole, especially when strong
backreaction is included. We can model this complete system by an
interacting quantum field. A particle- field system is a particular
case of it. Of course a black hole is different from a particle.
In this modeling, we will first explain how information is registered or `lost'
in an interacting quantum field, then the distinct features of a black hole
and finally, the information loss paradox of black hole systems.

To approach the black hole information loss paradox we need to understand
three conceptual points: \\
1) How does one characterize the information
content of a quantum field (interacting, as a model for the black hole -
quantum field closed system, with backreactions)\\
2) How does the information flow from one part of this closed system (hole)
to another (field) and vice versa in the lifetime of the black hole?
Does information really get `lost'? If yes, where has it gone?
Can it be retrieved? If no, where does it reside?\\
3) What is special about black hole radiation system as distinct from
ordinary particle / field system?\\

The following is a brief sketch of the picture we have developed
based on our studies of these issues in varying depths
in the past ten years. The development of the correlation dynamics in quantum
fields formalism was done with Esteban Calzetta \cite{cddn},
that of viewing black hole radiance and inflation as exponential scaling
was explored with Yuhong Zhang \cite{HuEdmonton,cgea,jr}
partly based on work on critical phenomena done earlier with
Denjoe O'Connor \cite{HuO'C}.
Studies of simple models to illuminate various points in this conceptual scheme
are being pursued now with Alpan Raval. For general background on this issue,
see the  review of Page \cite{Page} which contains a comprehensive list
of references till 1993, and Bekenstein \cite{Bekenstein}.
For string theory related ideas, see, e.g., the recent reviews of
\cite{bhstring}. Our approach is closer in spirit to \cite{Wilczek,AngRecoh}.

\subsection{Correlation functions as registrar  and
correlation dynamics as flow-meter of information in quantum fields}

The set of correlation functions provides us with the means to register
the information content of a quantum field. As mentioned above, the
complete set of $ n = \infty $ correlation (Wightman) functions
carries the complete information about  the quantum state of the field.
A subset of it which defines the subsystem, such as the mean field and the
2-point function, as is used in the ordinary description of (effective)
field theory,  carries only partial information. The missing information
resides in the correlation noise, and manifests as dissipation in the
subsystem dynamics. In this framework, the entropy of an incompletely
determined quantum system is simply
given by $ S = - Tr \rho_{red} ln \rho _{red}$, where
the reduced density matrix of the subsystem (say,  consisting of the
lower correlation orders) is formed by integrating out the environmental
variables (the higher correlation orders) after the hierarchy is
truncated with factorization conditions.

While the set of correlation functions act as a registrar  of information
of the quantum system, keeping track of how much information resides in
what order,
the dynamics of correlations as depicted by the hierarchy of equations
of motion derived from the master effective action depicts the flow of
information from one order to another, up or down or criss-crossing the
hierarchy. Correlation dynamics has been proposed for the
description of many body systems before \cite{Balescu}, and applied to
molecular and plasma kinetics.  In the light of the above theoretical
description we see this scheme as a potentially powerful way to do
quantum information systematics, i.e.,
keeping track of the content and flow of information
in a coherent or partially coherent quantum system.

\subsection{Information appears lost to subsystems of lower order
correlations -- `missing' information stored in higher order correlations}

Most measurements of a quantum field system are of a local or quasilocal
nature.
If one counts the information content of a system based on the mean field
and the lowest order correlation functions, as in the conventional
way (of defining quantum field theory in terms of, e.g., 2PI
effective action),  one would miss out a good portion of the information in
the complete system, as much of that now resides in the higher order
correlation functions in the hierarchy. These invoke nonlocal
properties of the field, which are not easily accessible in the ordinary
range of accuracy in measurements. Such an observer would then report on a loss
of
information in his way of accounting (which is taken to be in agreement
with other observers with the same level of accuracy of measurement).
Only observers which has access to all orders (the `master' in the
master effective action) would be able to see the complete development
of the system and be able to tell when the nth order observer begins to
lose track of the information count and reports an information loss.
This is more easily seen in molecular dynamics:
For observers confined to measuring one particle distribution
functions (truncation of the BBGKY hierarchy) and with the molecular chaos
assumptions implicitly invoked (factorization condition), he would report
on information loss. This is how Boltzmann reasons out the appearance of
dissipation in ordinary macroscopic physical phenomena.
The same can be said about measurement of quantum systems.

\subsection{Exponential scaling in Hawking effect facilitates information
transfer to the higher correlations. Black hole with its radiation
contains full information, but retrieval requires probing the higher order
nonlocal properties of the field}

How is this scheme useful in addressing the black hole information
problem?  How is the black hole / quantum field system different from
the ordinary cases? The above scheme can explain the apparent loss of
information in a quantum system, but there is an aspect distinct to
black holes or systems emitting thermal radiance.
Some years ago, I made the observation that all mechanisms
of emission of (coherent) thermal radiance such as
the Hawking effect in black holes, or the Unruh effect in accelerated
detectors,
involve an exponential redshifting process of or by the system.
This can be compared to the
scaling  transformation in treating critical phenomena.
After sufficient exponential redshifting
(at late times of collapse) and the black  hole is emitting thermal radiance,
the system has reached a state equivalent to the approach to a critical point
in phase transition. There, the physical properties of the
system are dominated by the infrared behavior, and
as such, the lowest order correlation functions are no longer sufficient to
characterize the critical phenomena. The contribution of higher order
correlation functions would become important.
% partake in shaping the nature and dynamics of the phase transition.
Note that in ordinary situations, only the
mean field and the 2 or 3  point correlation functions are needed to
give an adequate description of the dynamics of the system.
But for black holes or similar systems where
exponential red-shifting is at work, higher order correlations are readily
activated. The information content profile for a quantum field in the
presence of a black hole would be very different from ordinary systems,
in that it is more heavily populated in the higher  end of the spectrum
(of correlation orders). If one carries out measurement at the lower
end of the spectrum, one would erroneously conclude that there is information
loss.

So, following the correlation dynamics of the black hole / field system,
while the state of the combined system remains the
same as it had begun, there is a continuous shifting of information content
from the black hole to the higher correlations in the field as it evolves.
Correlation dynamics of fields can be used to keep track of this information
flow. We speculate that the
information content of the field will be seen to shift from low correlation
orders to the higher ones as Hawking radiation begins and continues. The
end state of the system would have a black hole evaporated, and its information
content transferred to the quantum field, with a significant portion of it
residing in the higher order nonlocal correlations.

This is, however, not the end of the story for the correlation dynamics
and information flow in the field. The
information contained in the field will continue, as it does in general
situations, to shift across the hierarchy.
As we know from the BBGKY description of
molecular dynamics, after  the higher correlation orders in the hierarchy
have been populated -- and for systems subjected to exponential red-shifting
this condition could be reached relatively quickly -- the information
will begin to trickle downwards in the hierarchy, though far slower than the
other direction initially. The time it takes (with many criss-crossing)
for the information to return to the original condition is the Poincare
recurrence
time. This time we suspect is the upper bound for the recoherence time,
the time for a coherent quantum system interacting with some environment
to regain its coherence  \cite{AngRecoh}.
It would be interesting to work out the information flow using the
correlation dynamics scheme for a few sample systems, both classical
and quantum,  so as to distinguish the competing effects of different
characteristic processes in these systems, some quantum, some statistical
(e.g., decoherence time, relaxation time, recoherence time and recurrence
time).

 Our depiction above uses the interacting field model. Simpler cases might
 show somewhat degenerate behavior.
\footnote{An example is the interesting result of recoherence
reported by Anglin et al \cite{AngRecoh}.  We think their reported result of
a recoherence time of the order of the relaxation time is special to
the simple model of particle free-field interaction.
As the field modes couple only through their interaction with the particle,
and not amongst themselves, there is no structure or dynamics of the
information
content of the field itself, and the only time scale for it to return
is via interaction with the particle,
which is why  the recoherence time is related to the relaxation time of
the particle.  We expect in more general and complex systems (thus excluding
many spin systems) the recoherence time is much longer than the relaxation
time,
more in the order of the recurrence time.}
Details of these investigations will be reported in journal articles.\\

\noindent {\bf Acknowledgement}\\

I thank the director of this School, Prof. Norma Sanchez, for expertly
organizing a stimulating and enjoyable meeting.
My lectures were based on work I did in the last two years
with Esteban Calzetta, Andrew Matacz, Alpan Raval and Sukanya Sinha.
Research is supported in part by the National Science Foundation
under grant PHY94-21849.

\end{document}